\documentclass{PoS}

\usepackage{latexsym}
\usepackage{subfigure}
\usepackage{multicol}
\usepackage{multirow}
\usepackage{mathrsfs}
\usepackage{verbatim}
\usepackage{amsmath,amsthm, amssymb}

\newcommand{\stat}{\textrm{stat}\,}

\newcommand{\msbar}{\overline{\textrm{MS}}\,}

\title{Non-perturbative renormalization of the static quark theory in a large volume} 
\ShortTitle{Non-perturbative renormalization of the static quark theory}
\author{
	\speaker{Piotr Korcyl}\\
	Institut f\"ur Theoretische Physik, Universit\"at Regensburg, D-93040 Regensburg, Germany\\
        E-mail: \email{piotr.korcyl@ur.de}}
\author{Christoph Lehner\\
Physics Department, Brookhaven National Laboratory, Upton, NY 11973, USA\\
E-mail: \email{clehner@quark.phy.bnl.gov}}
\author{Tomomi Ishikawa\\
RIKEN-BNL Research Center, Brookhaven National Laboratory, Upton, NY 11973, USA\\
E-mail: \email{tomomi@quark.phy.bnl.gov}}

\definecolor{pink}{rgb}{1.0, 0.0, 1.0}

\abstract{
We report on progress to renormalize non-pertubatively
the static heavy quark theory on the lattice.
In particular, we present first results for position-space
renormalization scheme for heavy-light bilinears. We test our
approach on RBC's $16^3 \times 32$ lattice ensemble with
$m_{\pi} \approx 420$ MeV, Iwasaki gauge action and domain wall
light fermions.
}

\FullConference{The 33nd International Symposium on Lattice Field Theory\\
                 14-18 July, 2015\\
                 Kobe, Japan}

\begin{document}

\section{Introduction}

Studies of properties of heavy quarks and heavy mesons are an important part of lattice simulations
of Quantum Chromodynamics since it can be argued that they are especially sensitive to the physics beyond the Standard Model. 
Such calculations are however difficult to perform in a straigthforward manner because of 
a large hierarchy of scales, namely $m_B/m_{\pi} \approx 40$. Consequently it is useful to employ an effective 
description of heavy quarks, which in the case of the present contribution, is formalized as the Heavy Quark Effective Theory. 
More precisely we consider the leading order of HQET and we propose a new non-perturbative renormalization 
scheme, which allows for a fully non-perturbative renormalization of the action and of any composite operators of interest.
Our scheme is based on correlation functions in coordinate space and we discuss its practical implementation
for the static Lagrangian as well as for the temporal component of the heavy-light axial and vector currents. In the following 
we introduce very briefly the static quark theory (section \ref{section hqet}) and present details of our scheme 
(section \ref{section method}). Present numerical results for the ratio of renormalization constants 
$Z^{\textrm{stat}}_V/Z^{\textrm{stat}}_A$, heavy quark mass and the renormalization factors themselves 
are discussed in sections \ref{section mass}-\ref{section Z pert}. We conclude in Section \ref{section conclusions}.

\section{HQET: an effective field theory for QCD}
\label{section hqet}

HQET provides an effective description of QCD processes with initial and final states containing
a single heavy quark. The high momentum components of the massive quark field are integrated out and their
contribution is summarized in the HQET parameters, whereas the low momentum components are present as a 
new two-component effective field $\psi_h$. The quark mass dependence is explicitly removed and hence all 
masses computed within HQET must be shifted by an overall energy offset. At leading order, in the so-called 'static quark theory', 
one assumes that the heavy quark is infinitely heavy and the Lagrangian reduces to
\begin{equation}
\mathscr{L}_{\stat} = \bar{\psi}_h D_0 \psi_h.
\end{equation}

A peculiarity of this theory is that $D_0$ has the same quantum numbers as the lower dimensional mass operator 
and thus those two operators mix under renormalization with a linearly divergent coefficient. Note that Wilson 
fermions suffer from the same linearly 
divergent additive mass renormalization. In the latter case the non-perturbative renormalization is usually done in a hadronic 
scheme, where the mass of an appropriate meson is tunned to its experimental value.

It was argued that renormalization of the HQET Lagrangian must be done non-perturbatively otherwise uncancelled 
divergent terms can combine with lattice artefacts giving finite, non-vanishing contributions \cite{sachrajda nonpert}. 
A practical prescription of such renormalization for HQET at any order in $1/m$ was worked out by the ALPHA collaboration. It
consists in renormalizing HQET and matching it to QCD in a small volume using SF boundary conditions \cite{alpha}. 
Then the HQET parameters are evolved non-perturbatively using step scaling techniques to a large volume where 
hadronic matrix elements can be evaluated. At the leading order of HQET one is allowed to use perturbation theory \cite{tomomi2, tomomi}, 
however at the cost of large systematic errors. In this work we discuss a non-perturbative renormalization scheme which 
avoids using correlation functions with SF boundary conditions. 

\section{Description of the method}
\label{section method}

Our primary object of interest is the correlator of two heavy-light currents
\begin{equation}
C_{\Gamma}(t) = \langle \bar{\psi}_h(t, \vec{0}) \Gamma \psi(t, \vec{0}) \bar{\psi}(0, \vec{0}) \Gamma \psi_h(0, \vec{0}) \rangle.
\label{correlator}
\end{equation}
It depends only on the time separation, since the heavy quark does not propagate in space.
The correlator Eq.\eqref{correlator} contains all the information we need. The rate of decrease of $C_{\Gamma}(t)$ encodes 
the energy shift, and hence the large distance behavior of $C_{\Gamma}(t)$, where the slope corresponds 
to the static energy of the heavy-light meson, can be used to define a renormalization condition for the
static quark mass. The slope denoted by $\delta m(t^*)$ can be estimated in the following way \cite{sachrajda}
\begin{equation}
\delta m(t^*) = - \log \Big( \frac{C_{\Gamma}(t^*+1)}{C_{\Gamma}(t^*)} \Big). 
\end{equation}
Hence, on each configuration we extract the heavy-light meson mass and define the mass-renormalized correlator $\tilde{C}_{\Gamma}(t)$
\begin{equation}
\tilde{C}_{\Gamma}(t, t^*) = e^{-\delta_m(t^*) (t+1)} C_{\Gamma}(t).
\label{mass renormalized}
\end{equation}
An additional term $e^{-\delta_m(t^*)}$ was explicitly added in Eq.\eqref{mass renormalized}.
Such factor is commonly included in studies using off-shell renormalization conditions and reabsorbed into the 
heavy quark wave function renormalization (see for example \cite{tomomi,tomomi2}). We follow this convention. 
Note, however, that our scheme is an on-shell 
scheme and is defined using gauge-invariant quantities and hence the wave function 
renormalization factor does not appear explicitly. In other words our set of renormalization conditions 
is consistent with or without that factor. 
Note also that a renormalization condition Eq.\eqref{mass renormalized} imposed at large $t^*$ would be inaccessible in 
perturbation theory and hence would prohibit the translation to the usually employed continuum 
$\overline{\textrm{MS}}$ scheme. If one works in this non-perturbative regime of $t^*$ a solution would be to perform 
a non-perturbative step scaling to a higher scale, where contact to perturbation theory can be made. 
Otherwise, if the lattice spacing is fine enough one can choose $t^*$ small so that perturbation theory is reliable at that scale.

Subsequently, we use the correlator $\tilde{C}_{\Gamma}(t,t^*)$ to impose the renormalization condition which fixes the renormalization 
constant of the heavy-light current
\begin{equation}
\big(Z^{\textrm{X}}_{\Gamma}(t_0)\big)^2 \tilde{C}_{\Gamma}(t_0,t^*)  = C^{\textrm{lattice tree-level}}_{\Gamma}(t_0).
\label{ren condition}
\end{equation}
We stress that one can considerably suppress discretization effects by using the lattice free correlator on the right 
hand side of this equation. We now describe how the above conditions work in practice.

\section{Results}

\subsection{Ensemble details}

In this feasibility study we used one ensemble generated by the RBC collaboration
with three flavors of domain-wall light fermions on Iwasaki gauge action \cite{ensemble}. The lattice was $16^3 \times 32$ with a lattice spacing
$a = 0.11 \textrm{fm} = 1.73 \textrm{GeV}^{-1}$. The pion mass was estimated on this ensemble to be $m_{\pi} \approx 420$ MeV.
The following results were obtained using 20 configurations separated by 200 MDU so that autocorrelations 
in the HMC time are negligible. The analysis was done with two sets of propagators: point-to-all propagators and stochastic 
random $Z_2$ wall propagators. The correlation functions were averaged over the spatial volume at the sink. 
We used three definitions of the lattice static theory: the original Eichten-Hill proposal 
\cite{eichten}, which will be denoted in the following by EH, and two smeared versions proposed in \cite{sommer} denoted 
by HYP1 and HYP2. We observed (see figure \ref{axial Z}) that in general it is advantageous to use stochastic wall sources 
instead of point sources because the statistical error is significantly smaller. Also, the smeared actions behaved
better at larger distances than the Eichten-Hill action. Besides the unitary quark mass $am = 0.01$ we used two other values: 
one heavier and one lighter by a factor 2.


\subsection{Mass renormalization}
\label{section mass}

We use mass renormalization condition as explained in Eq.\eqref{mass renormalized} for $t^* = 8$ where the statistical noise
is under control. Note that according to \cite{tomomi} the effective mass plateau starts at $t=10$. As 
will be shown in section \ref{section Z pert} the lattice spacing of our ensemble is too coarse to 
reliably make contact with perturbation theory and therefore in this feasibility study we take the perspective that a step scaling 
calculation will be needed to translate our results to the $\msbar$ scheme. Hence, at this stage of our investigation we decided 
to renormalize  the static quark mass with $t^*$ in this non-perturbative regime.
We checked that the estimated mean values of $\delta m(t^*)$ agree with the results published in Ref.\cite{tomomi} for the HYP1 and HYP2
static actions. The mass renormalization condition was imposed on the $C_{\gamma_5 \gamma_0}(t)$  correlator at the unitary 
mass $am=0.01$ configuration per configuration and used for all remaining correlators. 


\subsection{$Z_A^{\stat}$ and $Z_V^{\stat}$ in position space scheme}
\label{section Z}

\begin{figure}
\begin{center}
\includegraphics[width=0.45\textwidth]{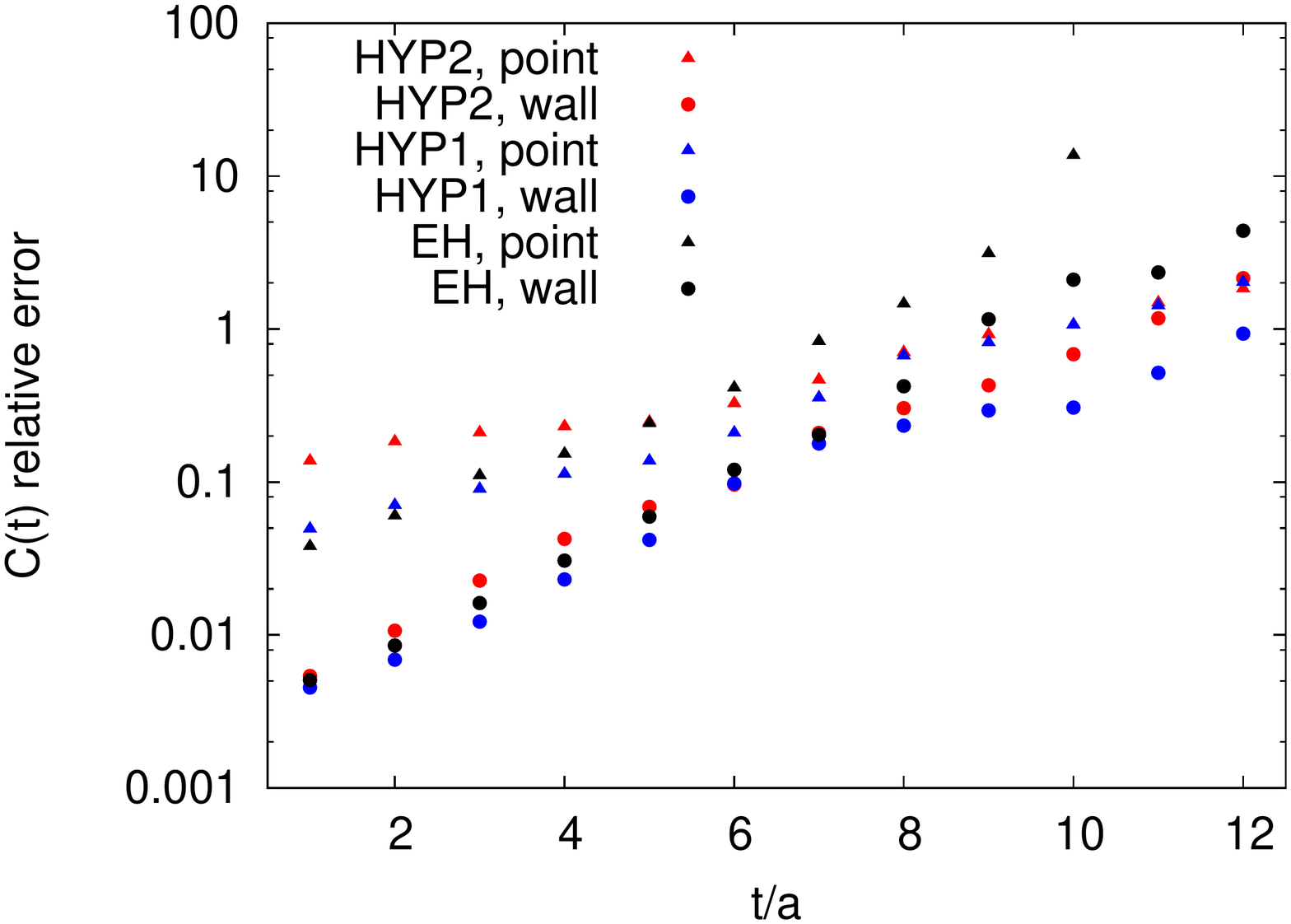}
\includegraphics[width=0.45\textwidth]{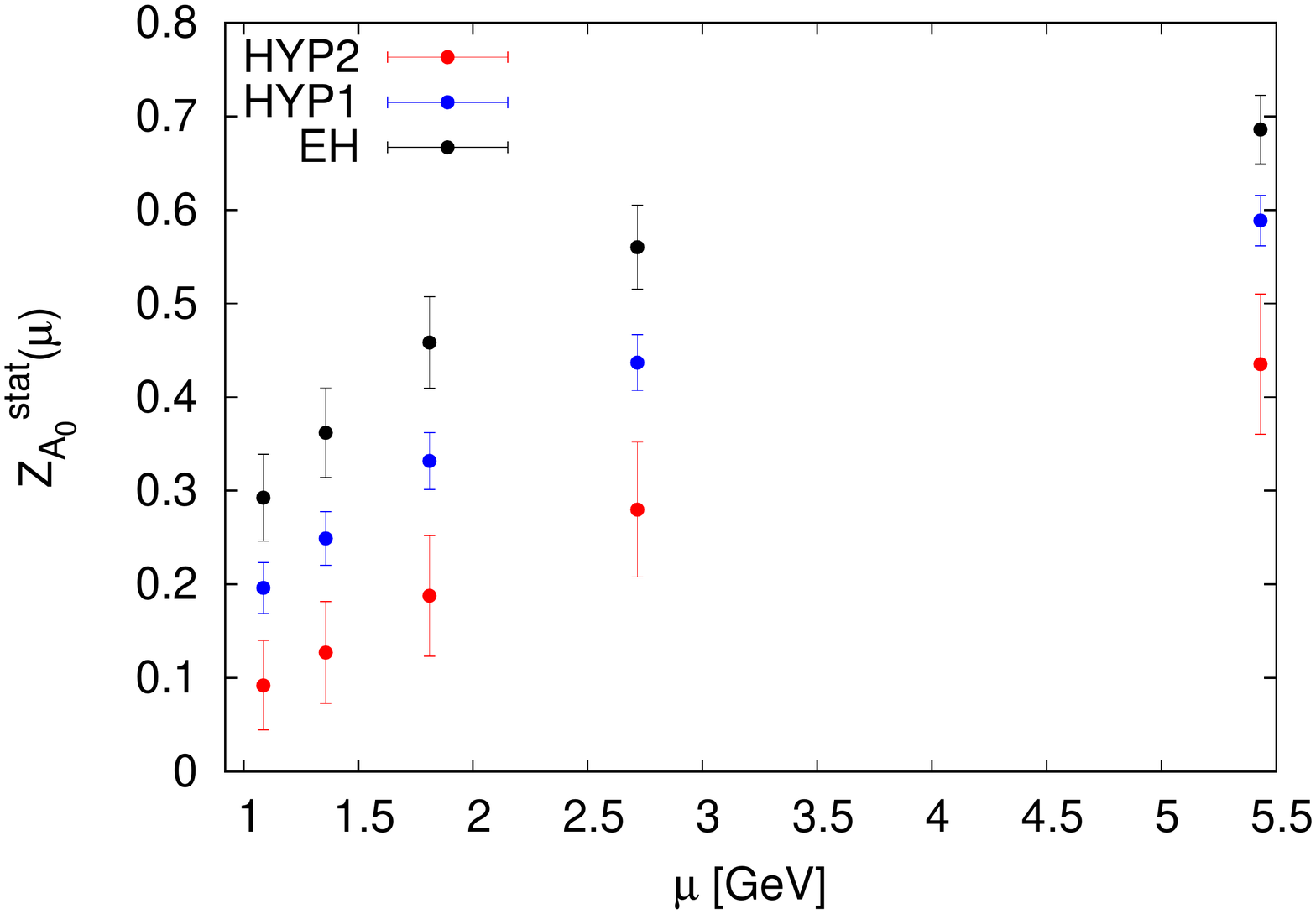}
\vspace{-1cm}
\caption{Left panel: Relative errors of the correlator $C(t)$ estimated from point and wall sources. Right panel: $Z^{\stat}_{A_0}$ in the chiral limit. Different data sets correspond to different static actions. The scale $\mu$ corresponds to $\mu=\pi/t_0$.\label{axial Z}}
\end{center}
\end{figure}

The currents' renormalization constants $Z_A^{\stat}$ and $Z_V^{\stat}$ were obtained by imposing
the condition \eqref{ren condition} on the correlation function $\tilde{C}_{\Gamma}(t)$. The values
computed at $am=0.005, 0.01$ and $am=0.02$ were extrapolated linearly to the chiral limit. 
The results are shown on the right panel of figure \ref{axial Z}.

\subsection{$Z_{A}^{\stat}/Z_V^{\stat}$ as a check of precision}
\label{section ratio}

In order to test the entire setup we computed the ratio of the renormalization constants
\begin{equation}
R(t_0/a,am)=Z_{A}^{\stat}(t_0/a,am)/Z_V^{\stat}(t_0/a,am),
\end{equation}
which is independent of the mass renormalization. $R(t_0/a, am)$ is expected to be equal to unity in the high energy regime 
where effects of chiral symmetry breaking become negligible and therefore provides a check for our scheme. 
By using a chirally symmetric discretization of the Dirac operator one can bring this regime down so that it can be approached on 
current lattices. In order to check the above statement we performed a combined, correlated fit to all our data using the 
following fit ansatz
\begin{equation}
R(t_0/a, am) = R + \alpha / (t_0/a)^2 + \beta (t_0/a)^2 + \gamma am,
\label{eq fit}
\end{equation}
where the term proportional to $\beta$ is supposed to describe non-perturbative low-energy effects
and the term proportional to $\alpha$ parametrizes the discretization errors.
\begin{figure}
\begin{center}
\includegraphics[width=0.45\textwidth]{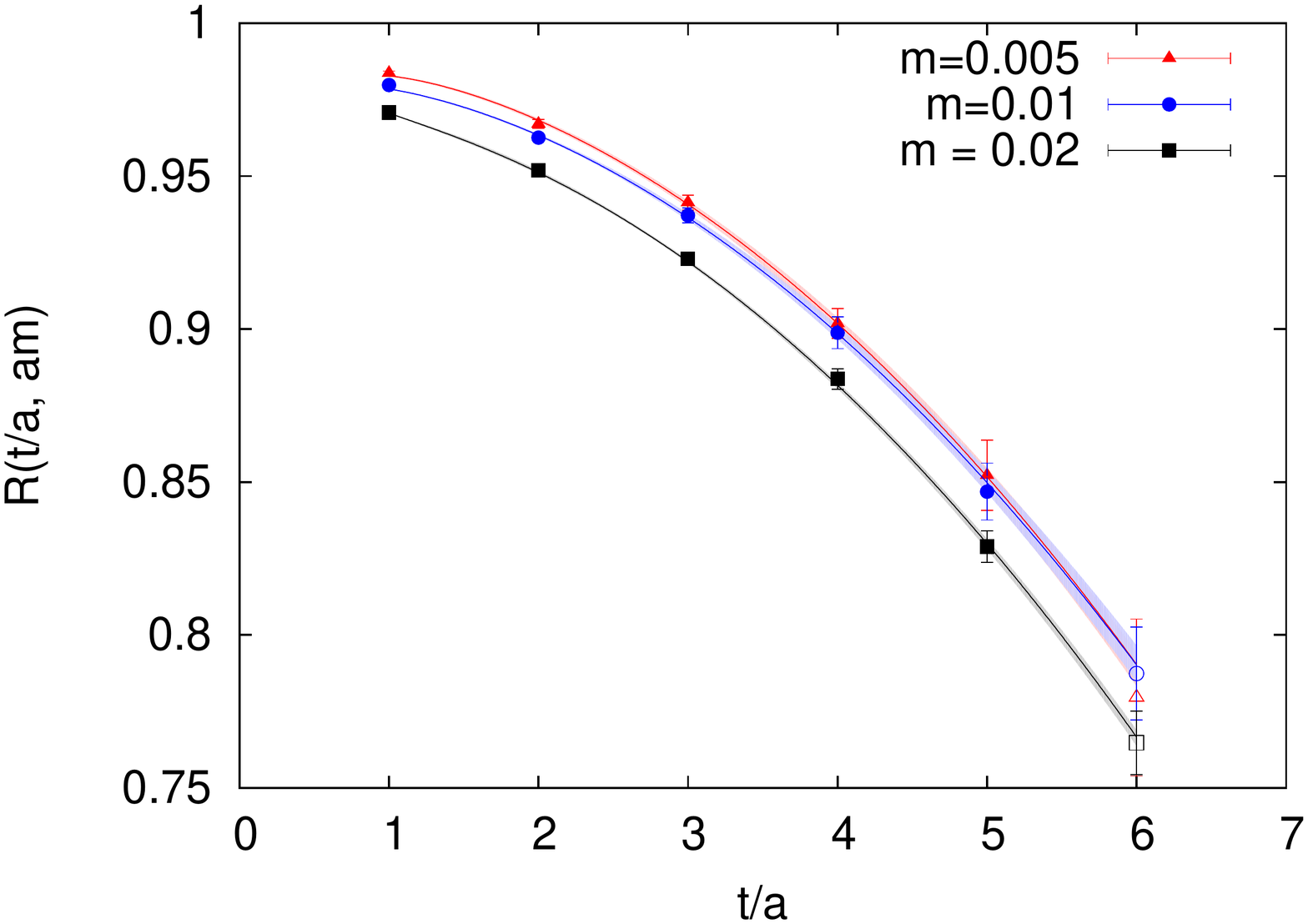}
\includegraphics[width=0.45\textwidth]{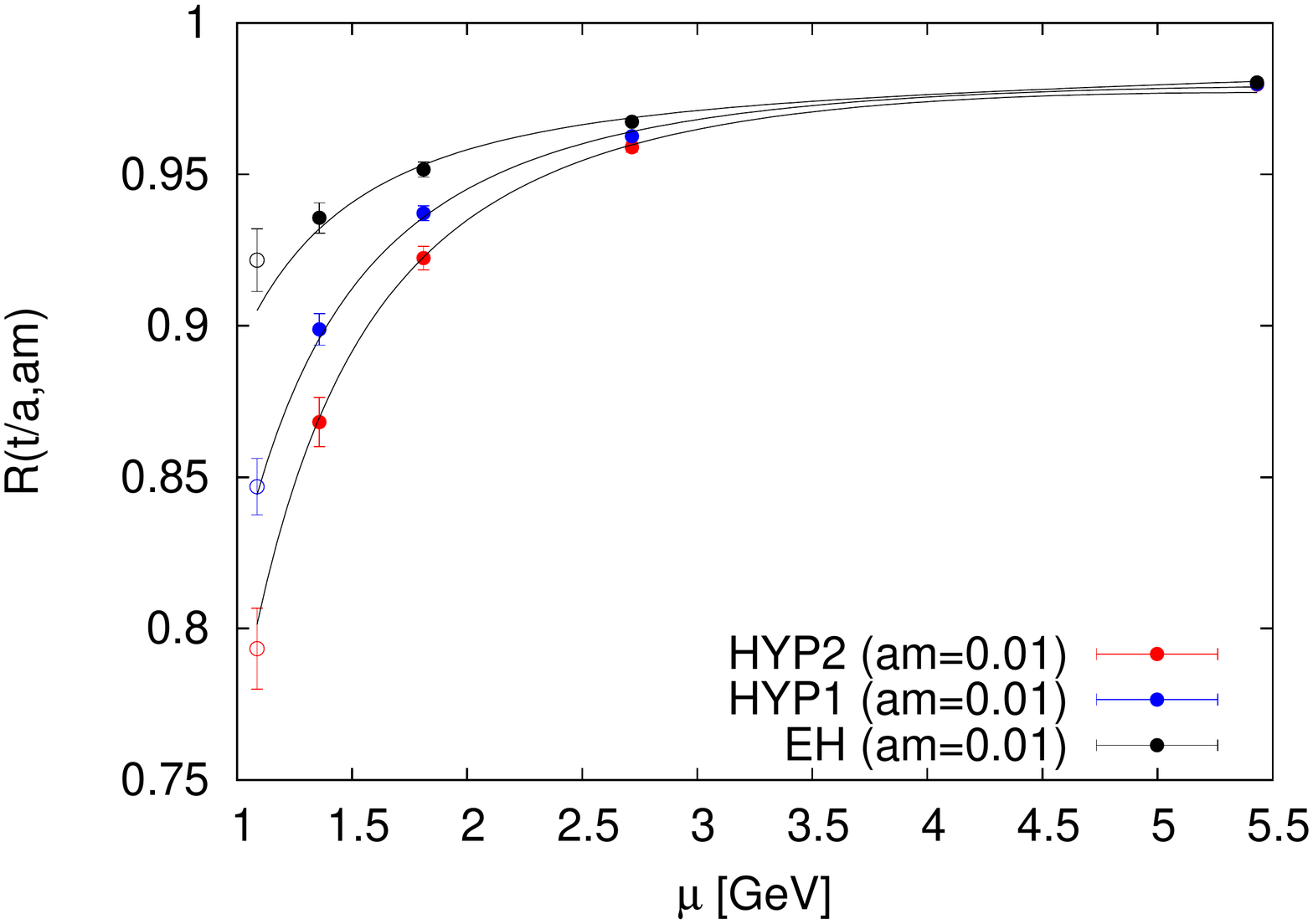}
\vspace{-1cm}
\caption{Left: Mass dependence of the ratio $R$ for HYP1. Right: Data for the ratio $R$ for EH, HYP1 and HYP2 action for am=0.01. \label{fig ratio}}
\end{center}
\end{figure}
The left panel of figure \ref{fig ratio} shows the dependence of $R$ on the quark mass. We see that $R$ approaches unity as $am$, 
the explicit chiral symmetry breaking term, is decreased. On the right panel we show the comparison of $R$ for different static actions. 
For the improved actions, HYP1 and HYP2, we find the extrapolated value to be equal to unity with a good precision. 
The calculation with the original Eichten-Hill proposal is one standard deviation away, but it is known to suffer from large discretization
effects. Results for the fit parameters are collected in table \ref{tab fit ratio}.
The first error is statistical and the second is an estimate of the systematic error obtained by the maximal difference between fit
results when changing the fitting range, i.e. including or excluding points with $t/a=5,6$. We conclude that our setup is working 
as expected and that for this ratio cutoff effects are under control even at small distances of few lattice spacings.
\begin{table}
\begin{center}
\begin{tabular}{|c|c|c|c|c|}
\hline
& R & $\alpha$ & $\beta$ & $\gamma$ \\
\hline
E-H & 0.9875(9)(97) & 0.0041(7)(422) & -0.0030(2)(5) & -0.79(4)(40) \\
HYP1 & 0.9995(14)(96) & -0.0033(12)(50) & -0.0057(2)(3) & -1.16(9)(37) \\
HYP2 & 1.0013(27)(38) & -0.0075(32)(40) & -0.0076(3)(2) & -0.92(12)(10) \\
\hline
\end{tabular}
\caption{Results for the fit parameters from Eq.5 . The first error is statistical and the second is systematic.\label{tab fit ratio}}
\end{center}
\end{table}

\subsection{$Z_A^{\stat}$ and $Z_V^{\stat}$ in $\msbar$}
\label{section Z pert}

As another check of our results we translated our renormalization factors to the $\msbar$ scheme and evolved
perturbatively to a common scale of $3$ GeV. The continuum perturbative approximation for $C_{\Gamma}$ can be found 
in \cite{braun}. 
We use a one-loop conversion factor between continuum HQET position space scheme and continuum HQET $\msbar$ scheme 
and two-loop running. The summary of our final numbers is given in table \ref{table4}. In the third column we list the
values of the renormalization constant in the $\msbar$ scheme at $3$ GeV, whereas in the last column we translate them
to the RGI scheme by cancelling the running with a two-loop perturbative approximation. Both, the results at $3$GeV and the RGI values,
should be scale independent, whereas our results show a 15\% difference between subsequent scales at which we evaluated our renormalization
constant. Taking into account that the size of the coupling constant, i.e. the lattice spacing of our ensemble, is rather large and
we use only a two-loop approximation of the running at the scale as low as $\approx 1.5$ GeV, we conclude that 
the size of the discrepancies agrees with our expectations. This behavior should be milder on a finer lattice.
\begin{figure}
\begin{center}
\includegraphics[width=0.45\textwidth]{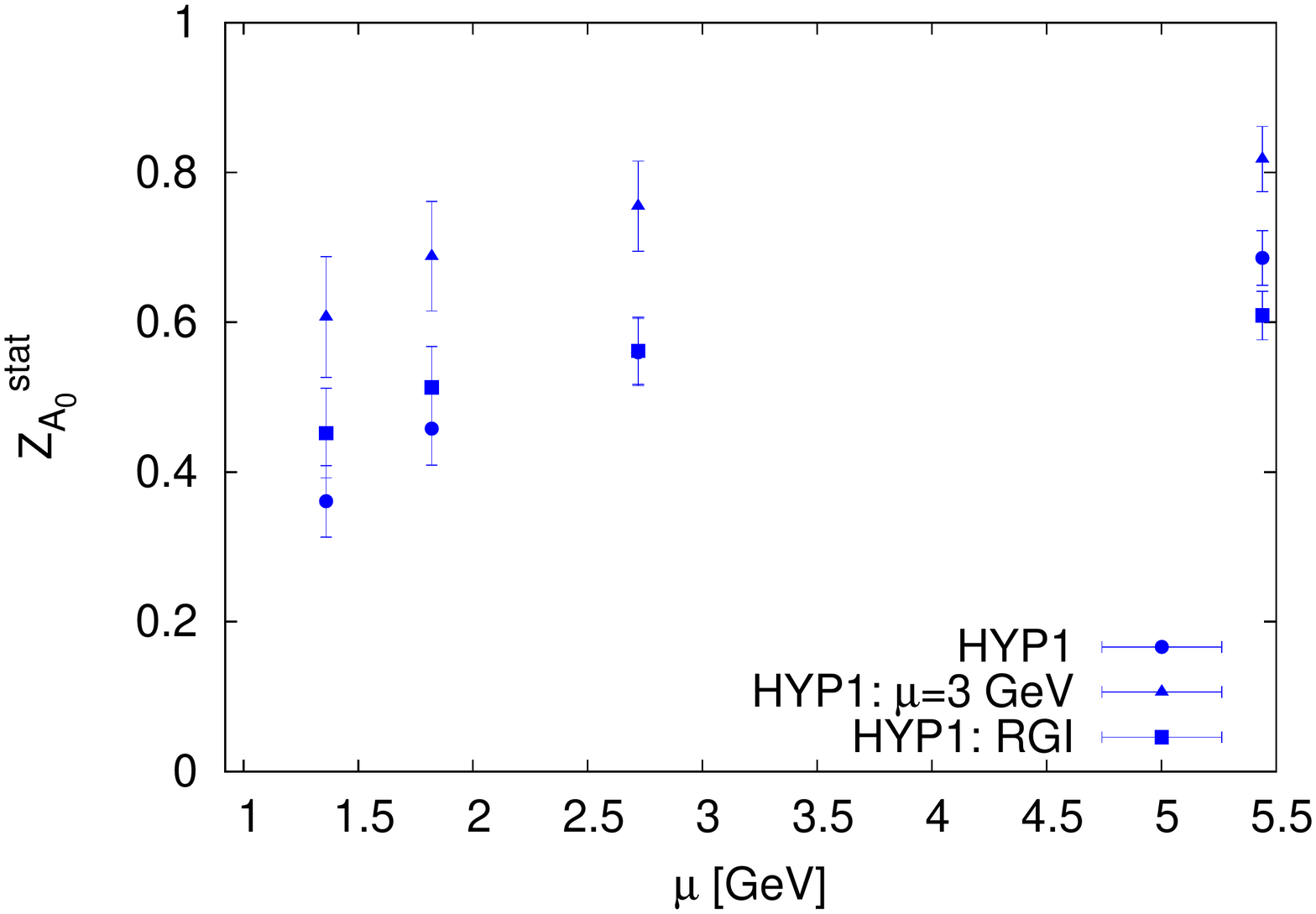}
\includegraphics[width=0.45\textwidth]{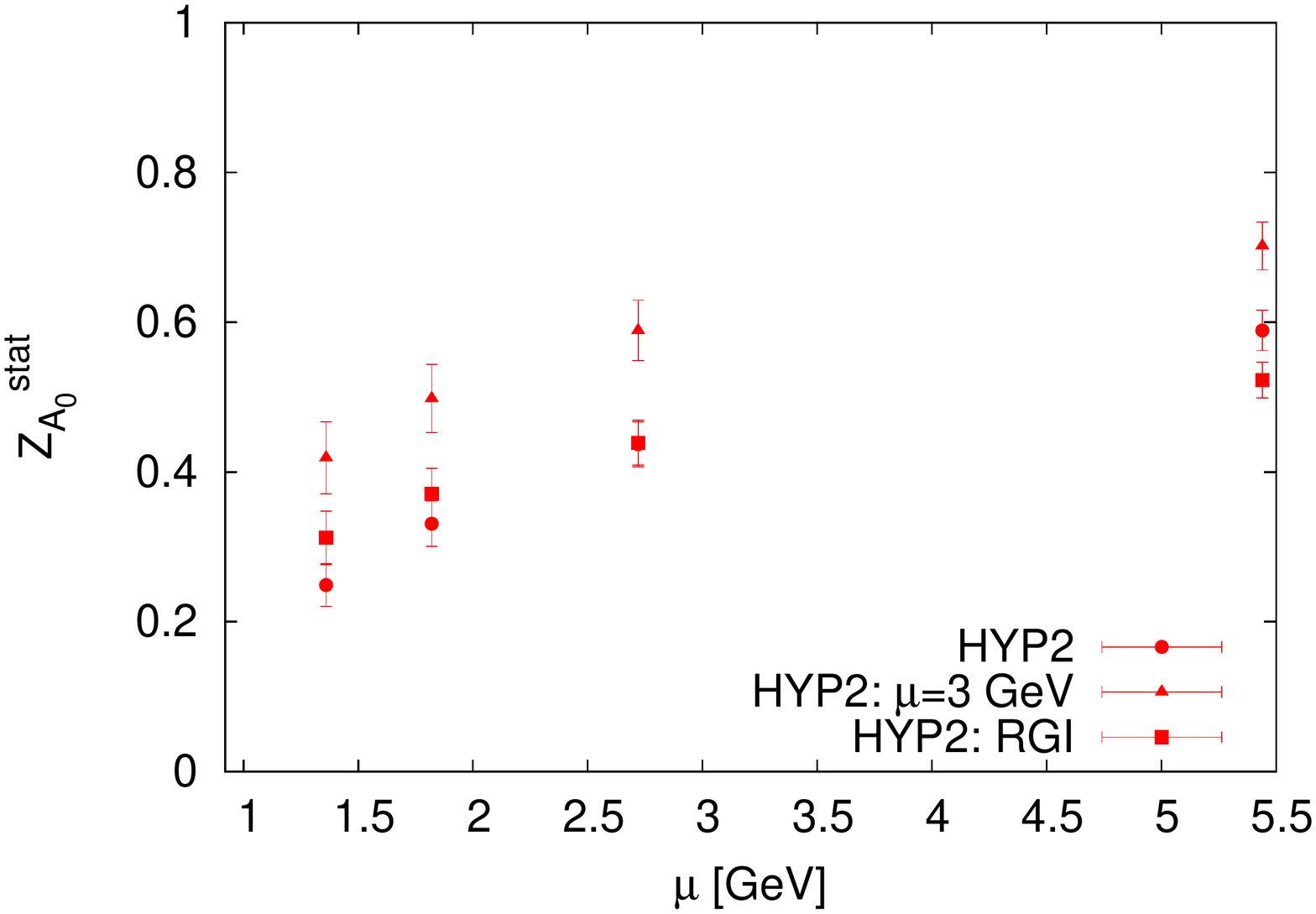}
\vspace{-0.5cm}
\caption{Comparison of $Z^{\stat}_{A_0}$, $Z^{\stat, \msbar}_{A_0}(3 \textrm{GeV})$ and $Z^{\stat, \msbar, \textrm{RGI}}_{A_0}$.}
\end{center}
\end{figure}
\begin{table}
\begin{center}
\begin{tabular}{|c|c|c|c|c|c|}
\hline
 &  & \multicolumn{2}{|c|}{HYP1} & \multicolumn{2}{|c|}{HYP2} \\
\hline
scale [GeV] & $\alpha_s$ & $Z^{\stat, \msbar}_{A_0}(3 \textrm{GeV})$ & $Z^{\stat, \msbar, \textrm{RGI}}_{A_0}$ &
$Z^{\stat, \msbar}_{A_0}(3 \textrm{GeV})$ & $Z^{\stat, \msbar, \textrm{RGI}}_{A_0}$ \\
\hline
5.44 & 0.1967 & 0.702(32) & 0.523(24) & 0.818(46) & 0.609(32) \\
2.72 & 0.2563 & 0.589(40) & 0.439(30) & 0.755(60) & 0.562(45) \\
1.81 & 0.3166 & 0.498(45) & 0.371(34) & 0.688(73) & 0.513(54) \\
1.36 & 0.3864 & 0.419(48) & 0.312(36) & 0.607(80) & 0.452(60) \\
\hline
\end{tabular}
\caption{Numerical values of renormalization constants in $\msbar$.\label{table4}}
\end{center}
\end{table}

\section{Discussion and conclusions}
\label{section conclusions}

In this contribution we proposed a non-perturbative renormalization scheme for the static quark theory
which can be used on large volume ensembles. The main ingredients of our proposal are renormalization conditions
formulated in position space, reduction of cut-off effects through a tree-level improvement and the use of
stochastic wall sources to decrease the statistical error of heavy-light correlators. The renormalization 
conditions are formulated using a single correlation function, which assures that they are gauge invariant and on-shell. 
Hence, the wave function renormalization constants are not needed. The main drawback of our renormalization scheme
is that it suffers from the same window problem as the well-known RI-MOM scheme. This shortcomming can be aleviated 
by using finer lattices or additional ensembles which then allow to run non-perturbatively the renormalization 
constants through the step scaling functions to higher scales where
the perturbative matching to $\msbar$ could be done more reliably.

In the future, we will provide a companion RI-MOM study and utilize a step-scaling setup.

\acknowledgments
P.K. would like to thank Norman Christ for many valuable discussions and for hospitality during his stay at Columbia University and 
acknowlegdes support of the Fulbright Commission through the Fulbright Senior Research Award. The computations were performed on the 
clusters at Columbia University and Fermilab as well as on the IDC cluster at the University of Regensburg.


\begin{thebibliography}{99}
\bibitem{sachrajda nonpert} C. Sachrajda, 
Nucl.Phys.Proc.Suppl. 185 (2008) 62-67
\bibitem{alpha} J. Heitger et al., 
JHEP 0402 (2004) 022, hep-lat/0310035 \\
B. Blossier et al., 
JHEP 1006 (2010) 002, arXiv:1001.4783 \\
B. Blossier et al., 
JHEP 1209 (2012) 132, arXiv:1203.6516
\bibitem{tomomi2} Y. Aoki, T. Ishikawa, T. Izubuchi, C. Lehner, A. Soni, 
Phys. Rev. D 91 (2015) 114505, arXiv:1406.6192
\bibitem{tomomi} T. Ishikawa, Y. Aoki, J. Flynn, T. Izubuchi, O. Loktik, 
JHEP 1105 (2011) 040, arXiv: 1101.1072
\bibitem{sachrajda} V. Gimenez, G. Martinelli, C. Sachrajda, M. Telavi, unpublished notes
\bibitem{ensemble} N.~H.~Christ et al.,  Phys.\ Rev.\ Lett.\ 105 (2010) 241601, arXiv:1002.2999 \\
 T.~Blum et al., Phys.\ Rev.\ D 84, (2011) 114503, arXiv:1106.2714
\bibitem{eichten} E. Eichten, B. Hill, Phys. Lett., B234, 511 \\
E. Eichten, B. Hill, Phys. Lett., B240, 193 \\
E. Eichten, B. Hill, Phys. Lett., B243, 427
\bibitem{sommer} A. Hasenfratz, F. Knechtli, Phys. Rev., D64, 034504 \\
M. Della Morte, A. Shindler, R. Sommer, JHEP 0508 (2005) 051, hep-lat/0506008
\bibitem{braun} E. Bagan, P. Ball, W.M. Braun, H.G. Dosch, Phys. Lett. B 278 (1992) 457
\end{thebibliography}
\end{document}